\documentstyle[aps,multicol,graphicx]{revtex}

\begin{document}
\title{Onset of dielectric modes at $110K$ and $60K$ due to local lattice
distortions in non-superconducting $YBa_{2}Cu_{3}O_{6.0}$ crystals}
\author{Z. Zhai, P. V. Parimi, J. B. Sokoloff and S. Sridhar}
\address{Physics Department, Northeastern University, 360 Huntington Avenue, Boston,\\
MA 02115.}
\author{A. Erb}
\address{DPMC, Universit\'{e} de Gen\`{e}ve, Gen\`{e}ve, Switzerland.}
\maketitle

\begin{abstract}
We report the observation of two dielectric transitions at $110K$ and $60K$
in the microwave response of non-superconducting $YBa_{2}Cu_{3}O_{6.0}$
crystals. The transitions are characterized by a change in polarizability
and presence of loss peaks, associated with overdamped dielectric modes. An
explanation is presented in terms of changes in polarizability of the apical 
$O$ atoms in the $Ba-O$ layer, affected by lattice softening at $110K,$ due
to change in buckling of the $Cu-O$ layer. The onset of another mode at 60K
strongly suggests an additional local lattice change at this temperature.
Thus microwave dielectric measurements are sensitive indicators of lattice
softening which may be relevant to superconductivity.
\end{abstract}

\begin{multicols}{2}%

It was recognized soon after the discovery of the high temperatures
superconductors that the cuprates are structurally similar to the
ferroelectric perovskites\cite{Bednorz}. The basic perovskite $ABO_{3}$
structure occurs in ferroelectrics like $BaTiO_{3}$ and incipient
ferroelectrics or quantum paraelectrics such as $SrTiO_{3}$, as well as in
sub-units of the superconductor $YBa_{2}Cu_{3}O_{6.0+x}$. The implications
of this structural similarity received early support from the observation of
large dielectric response \cite{Testardi92,Samara90} in the insulating
parent compound $YBa_{2}Cu_{3}O_{6.0}$. Furthermore, theoretical models have
been proposed which include the possible competition between
ferroelectricity and superconductivity \cite{Bishop89,Bussmann-Holder,Shenoy}%
.

In this paper we show some striking dielectric properties of single crystals
of insulating $YBa_{2}Cu_{3}O_{6.0}$ which seem to have a strong bearing on
the superconductivity of the doped YBCO. Traditionally the information on
lattice dynamics has been obtained from inelastic neutron, XAFS, Raman and
infrared spectroscopy measurements. Our microwave measurements probe
consequences of lattice modes on the long wavelength ($q=0$) dielectric
properties, and its very high sensitivity leads to the observation of
features not detected by other techniques.

In addition to large dielectric strengths $\varepsilon ^{\prime }\sim
10^{2}-10^{3}$, consistent with previous measurements, we report the
presence of two dielectric transitions at $110K$ and $60K$. These
transitions are accompanied by the onset of polarization modes indicated by
the presence of dielectric loss peaks below the transition temperatures. The
transitions arise from structural distortions occurring at these
temperatures, such as buckling of the $Cu-O$ plane leading to the $110K$
transition, which affect the electrodynamic response. Thus precision
microwave measurements are shown to be a sensitive probe of lattice effects,
complementing other traditional probes of lattice dynamics. Taken together
with numerous reports of lattice effects at or near the superconducting
transition temperature $T_{c}$ \cite{santafe92}, the present results
demonstrate the importance of charge and lattice dynamics in the high
temperature superconducting oxides.

Ultra-pure single crystals of $YBa_{2}Cu_{3}O_{6+x}$ were prepared in
contamination-free $BaZrO_{3}$ crucibles. The high quality of these single
crystals, which have also been prepared in the entire composition range from
insulating, non-superconducting material ($x=0$) to optimum ($x=0.95$) and
over-doped ($x=1.0$), has been extensively documented in a wide range of
measurements, including structural and transport studies of the
superconducting and non-superconducting states. A brief list of these
reports can be found in \cite{Erb99}. In this paper we focus on new results
on the insulating $x=0.0$ compound.

The high sensitivity microwave measurements were carried out in a $Nb$
superconducting cavity resonant at $10GHz$ in the $TE_{011}$ mode. The
sample is placed at the center of the cavity at a maximum of the microwave
magnetic field $H_{\omega }$. We introduce an electromagnetic susceptibility 
$\tilde{\zeta}_{H}(T)=\zeta _{H}^{\prime }(T)+i\zeta _{H}^{\prime \prime
}(T) $ which is related to the measured parameters, the shift in cavity
resonant frequency $\delta f(T)$ and the resonance width $\Delta f(T)$ by $%
\delta f(T)-i\Delta f(T)=-g(\zeta _{H}^{\prime }(T)+i\zeta _{H}^{\prime
\prime }(T)) $, where $g$ is a geometric factor. A detailed analysis of the
relevant cavity perturbation for general sample conditions including lossy
dielectric and metallic or superconducting states, has been recently carried
out by us \cite{zhairsi00}. We are able to directly measure the conductivity 
$\tilde{\sigma}_{tot}$ or the dielectric permittivity $\tilde{\varepsilon}%
_{tot}$ (where $\tilde{\sigma}_{tot}=-i\omega \varepsilon _{0}\tilde{%
\varepsilon}_{tot}$). The analysis shows that for arbitrary conductivity,

\begin{eqnarray}
\tilde{\zeta}_{H}(T) &=&-\frac{3}{2}\left[ 1-3/\tilde{z}^{2}+3\cot (\tilde{z}%
)/\tilde{z}\right] \text{ };  \label{dformula} \\
\text{where }\tilde{z} &=&ka=k_{0}a\sqrt{\tilde{\varepsilon}_{tot}}. 
\nonumber
\end{eqnarray}
Note that we use time dependences $e^{-i\omega t}$. In the limit $\tilde{z}%
\ll 1$, $\tilde{\zeta}_{H}(T)\approx (1/10)(k_{0}a)^{2}(\tilde{\varepsilon}%
_{tot}-1)$ for a lossy dielectric. The dielectric permittivity $\tilde{%
\varepsilon}_{tot}$ was extracted from the data using Eq. \ref{dformula}. $%
\tilde{\varepsilon}_{tot}$ includes both bound polarization ($\tilde{%
\varepsilon}$) and free charge conductivity $\tilde{\sigma}$ contributions,
i.e. $\tilde{\varepsilon}_{tot}=\tilde{\varepsilon}+i\tilde{\sigma}/\omega
\varepsilon _{0}$. In the present case the conductivity is negligible and $%
\tilde{\varepsilon}_{tot}=\tilde{\varepsilon}$. We have earlier carried out
extensive measurements of the surface impedance of a variety of
superconductors\cite{Srikanth96}, metals and insulators, and demonstrated
the validity of these measurements.

The dielectric permittivity $\varepsilon ^{\prime }(T)$ and $\varepsilon
^{\prime \prime }(T)$ of $YBa_{2}Cu_{3}O_{6.0}$ are shown in Fig. 1. Here $%
H_{\omega }||\hat{c}$-axis, so that the displacement currents are in the $ab$%
-plane, i.e. we are measuring in-plane dielectric permittivity $\tilde{%
\varepsilon}_{ab}$. The large microwave dielectric permittivity observed in
the present composition seems to be a characteristic of some perovskite
oxides. Such large dielectric strengths $\varepsilon ^{\prime }\sim
10^{2}-10^{3}$ in non-metallic insulating $YBa_{2}CuO_{6.0+x}$ were reported
by Rey, et. al., \cite{Testardi92} for ceramic samples which were quenched
to retain the oxygen homogeneity. It is worth remarking that the present
crystals are also quenched from high temperature and this may be an
important requirement for the observation of this effect. We have observed
similar response in the microwave dielectric permittivity of another $%
YBa_{2}Cu_{3}O_{6.0}$ crystal (Fig. 2, bottom panel) obtained from a
different batch, confirming the presence of the dielectric transitions
reported here.

The data in Fig.1 can be analyzed in terms of three dielectric modes, $%
\tilde{\varepsilon}=\tilde{\varepsilon}_{\alpha }+\tilde{\varepsilon}_{\beta
}+\tilde{\varepsilon}_{\gamma }$, each of which is well described by a Debye
relaxation form with respect to the temperature dependence : 
\begin{equation}
\tilde{\varepsilon}=\tilde{\varepsilon}_{\alpha }+\tilde{\varepsilon}_{\beta
}+\tilde{\varepsilon}_{\gamma }=\sum_{i=\alpha ,\beta ,\gamma }\frac{%
\varepsilon _{i0}(T)}{1-i\omega \tau _{i}(T)}
\end{equation}

$\tilde{\varepsilon}_{\gamma }$ appears to represent the low $T$ tail of a
high temperature process, with $\varepsilon _{\gamma 0}(T)=160$, and with a
relaxation time $\tau _{\gamma }(T)=6.5\times 10^{-13}\sec ^{-1}\exp
(1000/T) $ characterized by an activation energy $1000K$. The $\tilde{%
\varepsilon}_{\gamma }$ process is dominant between $300K$ and approximately 
$180K$, below which it ``freezes'' out quasistatically as the dipole
relaxation rate becomes extremely slow. A residual temperature independent
dielectric contribution $\approx 465+i125$ remains at all temperatures. We
believe $\tilde{\varepsilon}_{\gamma }$ is the contribution which has been
measured by several previous investigators \cite{Testardi92,Samara90} on
non-metallic $YBa_{2}CuO_{6.0}$ and represents a polarization mode formed at
high temperatures $T>300K$. $\tilde{\varepsilon}_{\alpha }$ and $\tilde{%
\varepsilon}_{\beta }$ indicate the onset of two new dielectric modes which
turn on below transition temperatures $T_{c\alpha }=60K$ and $T_{c\beta
}=110K$. We describe these modes with the following parameters :

\begin{itemize}
\item  $\varepsilon _{\beta 0}(T)=60(1-(T/T_{c\beta }))$, $T_{c\beta }=110K$
and $\tau _{\beta }(T)=4\times 10^{-10}(\sec \cdot K)/(T+200)$ , for the $%
\tilde{\varepsilon}_{\beta }$ process, and

\item  $\varepsilon _{\alpha 0}(T)=280(1-(T/T_{c\alpha }))$, $T_{c\alpha
}=60K$ and $\tau _{\alpha }(T)=2.5\times 10^{-10}(\sec \cdot K)/(T+5)$, for
the $\tilde{\varepsilon}_{\alpha }$ process.
\end{itemize}

$\varepsilon _{\alpha 0}$ and $\varepsilon _{\beta 0}$ are similar to order
parameters which grow at temperatures below a transition. As $T$ is lowered,
both $\varepsilon _{\alpha }^{\prime }(T)$ and $\varepsilon _{\alpha
}^{\prime \prime }(T)$ increase initially due to the growing polarization.
However below a characteristic temperature both $\varepsilon _{\alpha
}^{\prime }(T)$ and $\varepsilon _{\alpha }^{\prime \prime }(T)$ begin to
decrease because the dipoles are no longer able to follow the microwave
field. The peak temperature $T_{p\alpha }$ $\sim 25K$ is determined by the
condition $\omega \tau _{\alpha }(T_{p\alpha })=1$ , although the peak for $%
\varepsilon ^{\prime }$ is at a higher $T$ than for $\varepsilon ^{\prime
\prime }$. The peaks are so-called dielectric loss peaks. Identical
arguments hold for $\tilde{\varepsilon}_{\beta }(T)$ also. Here the peak is
much broader and occurs at $T_{p\beta }\sim 75K$.

For the $\alpha $ and $\beta $ processes, the temperature dependence is too
broad to be described by an activated relaxation rate. We have found that a
relaxation rate which is linear in $T$, i.e. $\tau _{\alpha ,\beta
}^{-1}(T)\approx a_{\alpha ,\beta }\,(T+T_{(\alpha ,\beta )0})$, describes
the data very well as seen in Fig. 1, with $a_{\alpha }\approx 0.4\times
10^{10}(sec.K)^{-1}$, $T_{\alpha 0}=5K$ and $a_{\beta }\approx 0.25\times
10^{10}(sec.K)^{-1}$ and $T_{\beta 0}=200K$. Such relaxation rates with
linear $T$ dependences are well known in the copper oxide superconductors.

We note that similar large dielectric constants have been observed in other
copper oxides. In $Bi_{2}Sr_{2}(Dy,Y,Er)Cu_{2}O_{8}$, the parent compound of 
$Bi:2212$, large in plane $\varepsilon ^{\prime }\sim 10^{3}-10^{5}$ were
reported \cite{BSCCOeps}. It is also important to note that the dielectric
modes discussed here bear a strong similarity to the numerous modes observed
in the dielectric response of the perovskite $SrTiO_{3}$ \cite{Ang} at $%
65K,37K$ and $16K$. The dielectric loss peaks reported here are similar to
those observed in $La_{5/3}Sr_{1/3}NiO_{4}$ \cite{Hakim99} and $%
Sr_{14}Cu_{24}O_{41}$\cite{Zhai99}.

The present results indicate that at $110K$ \cite{100ktransition} and $60K$,
two new polarization onset transitions occur in $YBa_{2}CuO_{6.0}$. We note
that the $110K$\ and $60K$\ onsets cannot arise from any contamination of
the sample by a superconducting phase, since then the contribution should be
diamagnetic (negative $\varepsilon ^{\prime }$), opposite to what is
observed.

A scenario leading to such dielectric transitions can be arrived at starting
with the so-called Bilz model for ferroelectricity\cite{Bilz87}, which is
based upon the nonlinear polarizability of oxygen, and originally developed
for perovskite structures. This applies to a displacive type ferroelectric
where dipole moments are induced during the phase transition so that soft
mode concept becomes important. Above the displacive ferroelectric
transition, the oxygen atoms are essentially oscillating in a potential
well. Below the ferroelectric transition, a double-well is formed and the
oxygen atom then locks into one of the minima - the displacement then leads
to a large permanent polarization. In the present case the ferroelectricity
is prevented from occurring, either due to quantum fluctuations, as was
proposed for $SrTiO_{3}$ \cite{Mullerpara}, or due to coupling between Ba-O
and Cu-O layers(Fig. 3) \cite{Shenoy}. Consequently the $O$ potential is
greatly softened leading to the large dielectric permittivities observed.

We use a modification of the Bilz model specifically for the oxo-cuprate
superconductors implemented by Shenoy, et. al. \cite{Shenoy}. The
equation-of-motion of the Oxygen relative ion-electron coordinate $\vec{w}$
is given by $m_{e}(\stackrel{_{\cdot \cdot }}{\vec{w}}+\Gamma \stackrel{.}{%
\vec{w})}+D\vec{w}=Ze\vec{E}e^{-i\omega t}$, where $D$ is the SCPA curvature
of the anharmonic Oxygen electronic potential. For a driving electric field $%
\vec{E}e^{-i\omega t}$, the susceptibility $\alpha =Zew/E=Ze/m_{e}D(\omega
_{o}^{2}-\omega ^{2}-i\omega \Gamma )$. The dielectric constant $\varepsilon
=n\alpha /\varepsilon _{0}$ then becomes

\begin{equation}
\tilde{\varepsilon}=\frac{\varepsilon (0)}{(1-\omega ^{2}/\omega
_{0}^{2})-i\omega \tau }
\end{equation}
Here $\varepsilon (0)=nZe^{2}/D$ and $\tau =\Gamma /\omega _{0}^{2}$.

Rather large dielectric constants are feasible for soft modes. For the case
of the $O$ atom in the $Ba-O$ layer in $YBa_{2}CuO_{6.0}$ we have $%
n=1.2\times 10^{28}m^{-3}$. With $Z=1,\varepsilon _{0}=8.85\times 10^{-12}F/m
$, and using $\omega _{0}=2\pi f_{0}$, $f_{0}=3\times 10^{13}Hz\,$so that $%
D=3.3\times 10^{-2}N/m$, we get $\varepsilon (0)\sim 10^{3}$ comparable to
the experimental results. Note that despite the softening, the condition $%
\omega \ll \omega _{0}$ is well satisfied ($f=10^{10}Hz$), and the so-called
Drude-Lorentz form of a resonant mode given by Eq. 3 reduces to the Debye-
like relaxation forms used in Eq. 2. The above estimate indicates
considerable softening of the anharmonic $O$ potential. Indeed this can
happen because the curvature is extremely sensitive to interatomic forces in
these materials. A mechanism for such softening has been given by Shenoy,
et. al. for the layered HTS.

There have been extensive studies of lattice dynamics in superconducting $%
YBa_{2}Cu_{3}O_{6+x}$\cite{egami98}. One of the key features that has
emerged is that there are small structural distortions which occur although
there is no change in the overall structure. Particularly well-established
are the structural distortions reported at $T_{c}$ in $YBCO$, $Hg:1201$ and $%
Tl:2212$\cite{Sharma}. In $YBa_{2}Cu_{3}O_{6+x}$ the coupling between apical
O(1) and planar O(2) oxygen changes at the superconducting transition due to
the displacements of O(2) in the directions perpendicular to the CuO2 plane
(Fig. 3) changing the buckling in the plane \cite{egami98}. The nearness of
the $\beta $ dielectric mode around $110K$ to the $T_{c}=93K$ ($%
YBa_{2}Cu_{3}O_{6.95}$) is striking and suggests a possible change in the
O(2) dynamics in $YBa_{2}Cu_{3}O_{6.0}$ as well, and hence we attribute this 
$\beta $ mode to the change in the dynamics of O(2).

In ref. \cite{Shenoy} the possibility of dielectric modes in coupled Ba-O
and Cu-O layers is described. Including the various interatomic forces, it
can be shown that the curvature $D$ becomes a function of the buckling angle 
$\theta $. The change in buckling at $110K$ would lead to a change in the
mixing of the $ab$ plane acoustic and $c$-axis optic mode, resulting in a
new set of mixed modes which would move to lower frequency due to softening.
The resulting decrease in $D$ would then explain the $\beta $ mode at $110K$%
. The calcultion of Shenoy et al. \cite{Shenoy} based on the mean field
approach support our description of the dynamics of O(1).

The presence of the $60K$ mode suggests another local structural change at
this temperature scale, possibly in the chain layer. It is very interesting
to note that this temperature scale is present in the doped YBCO as well\cite
{zhai00}. In addition to the proximity of the $110K$ transition to the
optimum superconducting $T_{c}$ of $93K$ noted above, equally important is
the presence of a secondary temperature scale around $60K$ in certain
measurements of $YBa_{2}Cu_{3}O_{6+x}$\cite
{Srikanth96,berthier00,gagnon97,jung98}. This temperature scale has
manifested itself in various experiments whose connections are becoming
apparent only recently. Thus the present results may have important
implications for superconductivity in these materials.

In optimally doped single crystals of $YBa_{2}Cu_{3}O_{6.95}$ an additional
onset of pair conductivity at $65K$ was noted in ref.\cite{Srikanth96}, well
below the main $T_{c}=93K$. The consequences of this on the thermal
conductivity \cite{gagnon97} and vortex transport \cite{jung98}have been
observed. The present temperature scales also have striking resemblance to
transitions around $110K$ and $65K$ in over doped YBCO observed in NQR
measurements by Grevin et al. \cite{berthier00}, which have been interpreted
in terms of CDW correlations. In their results a short range CDW sets in the
Cu-O chains at $110K$ which becomes long range around $65K$. The formation
of CDW in the chains modulate the charge in the planes leading to a
transition between an inhomogeneous charge state to a low-temperature
ordered charge state in the planes. Keeping in view the fact that NQR probes
the electric field gradient around the Cu nucleus the NQR transitions could
as well be due to changes in local structure which leads to CDW formation in
the doped metallic YBCO. Together, the NQR, thermal conductivity and the
present results stress the importance of the $110K$ and $60K$ temperatures
scales. These results indicate that local distortions in the structure at $%
110K$ and $60K$ would lead to changes in polarization as we have observed in 
$YBa_{2}Cu_{3}O_{6.0}$ and charge ordering in doped YBCO. Microwave
conductivity measurements on doped crystals\cite{zhai00} from $x=0$ to $1$,
when taken together with these other observations, strongly suggest that
doping moves this onset temperature from $60K$ at $x=0$ to $\sim 70K$ at $%
x=1 $. At the superconducting $T_{c}$ increases with $x$ , but never exceeds 
$93K $. An interesting implication is that the implied structural distortion
transition at $110K$ may represent an upper limit on the superconducting
transition $T_{c}$.

In conclusion we have observed three paraelectric modes with temperature
onsets at $60K$, $110K$ and $>300K$ in the non-metallic $%
YBa_{2}Cu_{3}O_{6.0}.$ These modes are well described by the change in
polarizability of apical $O(1)$ oxygens due to change in lattice dynamics
with temperature. A striking observation is that the two low temperature
modes have direct connections with the transitions observed by traditional
lattice probes as well as NQR and thermal conductivity measurements on doped
superconducting $YBa_{2}Cu_{3}O_{6.0+x}$, strongly suggesting that oxygen
and lattice dynamics plays an important role in both the superconducting and
non-superconducting materials.

We thank C.Kusko, R.S.Markiewicz, C. Perry and S.R.Shenoy for useful
discussions.

This work was supported by ONR and NSF-ECS-9711910.

\narrowtext%

\begin{figure}
\includegraphics*[width=0.4\textwidth]{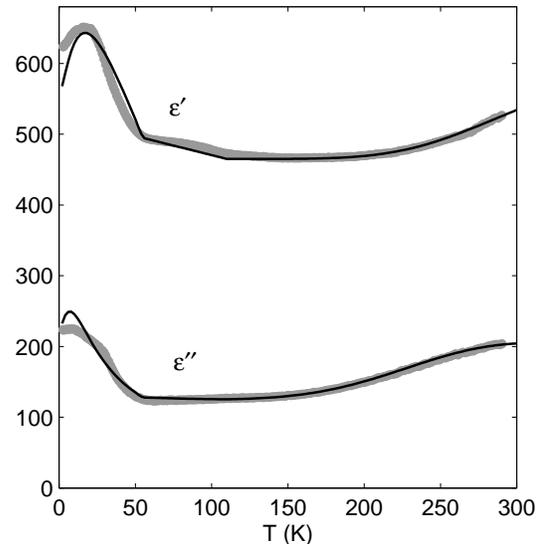}
  \caption{Dielectric constants $\varepsilon^{\prime}$ and  $\varepsilon ^{\prime \prime}$ at 10GHz of 
$YBa_2 Cu_3 O_{6.0+}$. Note the onset of dielectric response at $T_{cA} = 60K$  and 
$T_{cB} = 110K$. The solid lines represent $\varepsilon = \varepsilon_{\alpha} + \varepsilon_{\beta}
+ \varepsilon_{\gamma}$ as described in the text.}
\end{figure}
%

\begin{figure}
\includegraphics*[width=0.4\textwidth]{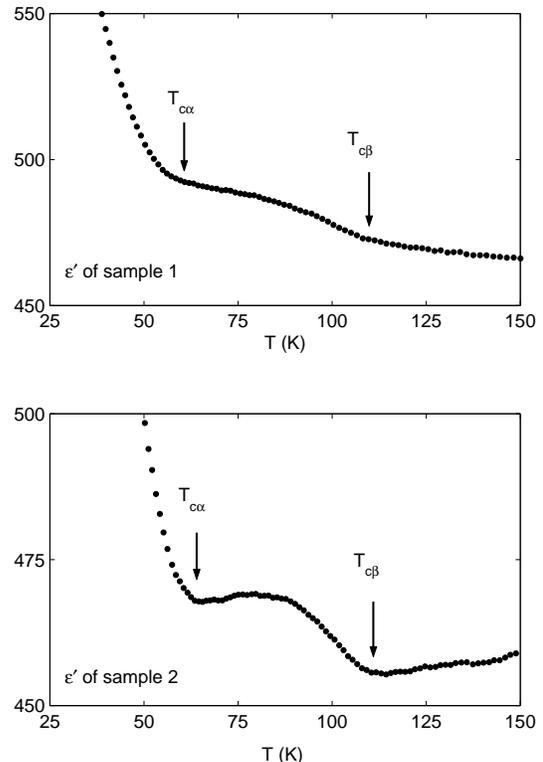}
  \caption{Dielectric constants $\varepsilon^{\prime}$ at 10GHz of $YBa_2 Cu_3 O_{6.0}$
 for samples 1 and 2. Note  the onset of
dielectric modes at $T_{c \alpha} = 60K$ and $T_{c \beta} = 110K$ are present
in both the smaples.}
\end{figure}
%

\begin{figure}
\includegraphics*[width=0.4\textwidth]{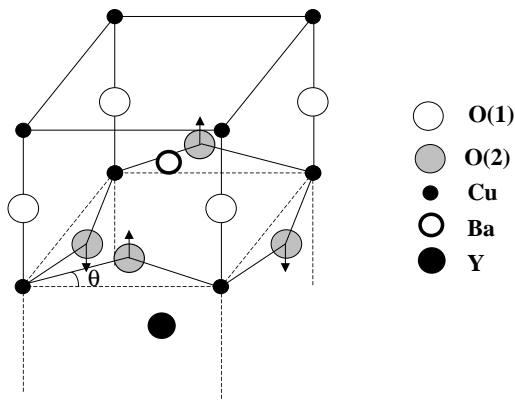}
 \caption{Apical, O(1) and planar O(2) oxygens in YBCO. 
The change in the buckling angle at 110K changes the polarizability of O(1) resulting in the $\beta$ dielectric mode. }
\end{figure} %

\end{multicols}%

\end{document}